# A Radiation-Balanced Silica Fiber Amplifier

Jennifer M. Knall,[1,*] Magnus Engholm,[2] Tommy Boilard,[3] Martin Bernier,[3] and Michel J. F. Digonnet[1]

[1]*Edward L. Ginzton Laboratory, Stanford University, Stanford, California 94305, USA*
[2]*Division of Electronics Design, Mid Sweden University, SE-85170 Sundsvall, Sweden*
[3]*Centre d'optique, photonique et laser (COPL), Université Laval, Québec, Canada QC G1V 0A6*

We report what we believe to be the first radiation-balanced fiber amplifier—a device that provides optical gain while experiencing no temperature rise. The gain medium is a silica fiber with a 21-μm-diameter core highly doped with $Yb^{3+}$ (2.52 wt.%) and co-doped with 2.00 wt.% Al to reduce concentration quenching. The amplifier was core-pumped with 1040-nm light to create anti-Stokes fluorescence (ASF) cooling and gain in the core at 1064 nm. Using a custom slow-light FBG sensor with mK resolution, temperature measurements were performed at multiple locations along the amplifier fiber. A 4.35-m fiber pumped with 2.62 W produced 17 dB of gain while the average fiber temperature remained slightly below room temperature. This advancement is a fundamental step toward the creation of ultra-stable lasers necessary to many applications, especially low-noise sensing and high-precision metrology.

Since the first rare-earth-doped fiber laser was fabricated using modified chemical vapor deposition (MCVD) in 1985 [1], tremendous advancements have been made in fiber laser technology. Fiber lasers now produce coherent emission ranging from the UV to the mid-IR, in continuous-wave, Q-switched, and mode-locked outputs. They can exhibit extremely stable single-frequency emission with linewidths as narrow as a few kilohertz [2], and they can be scaled up to 10 kW in a single-mode output [3], making them the brightest humanmade light sources. One of the main challenges limiting further power scaling and improvements in temporal and spatial quality is the internal heat introduced by the laser's quantum defect. The quantum defect is the energy difference between the pump and laser photons, and it is converted into heat through nonradiative relaxation. In $Yb^{3+}$, which has one of the smallest quantum defects, this energy difference is only 4% to 8% of the pump photon energy [4]. Yet this is enough that even a relatively low power 1-W Yb-doped laser can experience several degrees of heating [5]. Temperature variations induce instabilities in the laser frequency, which cause linewidth broadening and an increase in frequency noise [6]. In high-power lasers, they also limit the output power through the onset of transverse mode instability [7]. In the most extreme case, heating will fracture or melt the fiber.

Currently, thermal effects are mitigated by either water cooling or thermoelectric cooling. While these solutions are comparatively energy efficient, water-cooled systems are prone to leaking, add significant bulk to the laser, and induce vibrations that degrade the spectral and spatial beam quality. Thermo-electric coolers are generally vibration-free, but they tend to cool the fiber asymmetrically, which creates undesirable thermal gradients.

First proposed in 1999, heat mitigation through anti-Stokes fluorescence (ASF) cooling has emerged as a promising potential solution [8,9]. Cooling is induced when a gain medium is optically or electronically pumped at an energy lower than the average energy of the fluorescence [10]. Energy is extracted in three steps: (1) the pump excites electrons from a higher sub-level of the ground manifold of the laser ion to a low-lying sub-level of the upper manifold, (2) the electrons within the upper manifold annihilate phonons to redistribute themselves according to the Maxwell-Boltzmann distribution, thereby acquiring extra energy from the phonon bath, and (3) this extra energy (as well as the pump energy) is carried out of the sample when the electrons relax radiatively to the ground manifold and emit fluorescence with a higher average photon energy then the pump. Since the heat extracted per electron is the small energy difference between the pump and fluorescence photons, ASF cooling extracts relatively small amounts of energy per unit volume [10]. Therefore, any number of extraneous exothermic effects can either partially negate or overwhelm the heat extracted by ASF. The most common mechanisms are residual absorption of pump or fluorescence photons by impurities in the gain medium [11], and concentration quenching [12]. To achieve significant ASF cooling, it is paramount to select a combination of rare-earth ion and host composition that minimizes these two deleterious effects, especially the latter. Until recently, this limitation restricted ASF cooling to exotic crystals or fluorides with low quenching and absorptive loss, and experiments were often performed in a vacuum to minimize the heat load from air convection. This form of cooling adds minimal bulk to the laser, and no moving parts. It not only eliminates harmful vibrations, but also prolongs the lifetime of the laser, and minimizes maintenance time and cost.

The only ASF-cooled radiation-balanced laser reported to date used a highly doped Yb:YLF crystal [9]. The 3x120 mm rod was end-pumped with an array of 1030-nm Yb:silica fiber lasers, simultaneously inducing lasing at 1050 nm and ASF cooling. Since the extracted heat is proportional to the optically excited doped area, which is relatively large in a bulk laser, the extracted heat was significant, and the laser was able to output 80 W while maintaining zero average temperature change along the crystal. Since then, several theoretical papers have presented models for radiation-balanced operation in semiconductor [13] and fiber [14-16] lasers, and in fiber amplifiers [17,18], but none of these devices have been demonstrated experimentally. For fiber devices, this is partly due to the limited heat that can be removed from a fiber because of the small volume of the doped core [19]. Also, until recently, cooling in fibers had been primarily limited to fluorides [20-22], the only fiber host known to offer both high quenching-free rare-earth concentrations and low residual absorptive loss. Given the overwhelmingly commercial dominance of silica in fiber lasers and amplifiers, it was paramount to achieve cooling in this host.

Toward the end of 2019, this situation changed drastically with breakthrough work that demonstrated the first cooling of a Yb-doped silica fiber [23] and of a Yb-doped fiber preform [24]. A few months later, significantly more cooling was reported in a silica fiber with an improved composition [25], and in a silica preform placed in vacuum [26]. Temperature changes as large as -70 mK were measured in the fiber at atmospheric pressure [25], and up to 6 K of cooling was measured in the silica preform [26]. The work of [24-26] demonstrated that silica fibers with the right composition and fabrication method can be cooled with efficiencies close to those of fluoride fibers [22]. Capitalizing on this innovation, we report here the first internally cooled fiber amplifier.

In the limit of negligible absorptive loss, the maximum heat that can be extracted per unit length from a fiber doped with a two-level laser ion is quantified by [19]:

$$\left(\frac{dQ}{dt}\right)_{max} = \left(\frac{\tau_{rad}}{\tau(N_0)}h\nu_p - h\langle\nu_f\rangle\right)\frac{\sigma_p^a}{\sigma_p^a+\sigma_p^e}\frac{\pi a^2}{\tau_{rad}}N_0 \quad (1)$$

where $N_0$ is the ion concentration, $a$ is the radius of the doped core, $\tau_{rad}$ is the radiative lifetime of the ions, $h\nu_p$ is the energy of the pump photon, $h\langle\nu_f\rangle$ is the average energy of the fluorescence photons, and $\sigma_p^a$ and $\sigma_p^e$ are the pump absorption and emission cross-sections of the ions. The lifetime $\tau(N_0)$ is the total concentration-dependent upper-state lifetime, including quenching-induced relaxation. It depends on the degree of concentration quenching according to [12]:

$$\tau(N_0) = \frac{\tau_0}{1+\frac{9}{2\pi}\left(\frac{N_0}{N_c}\right)^2} \quad (2)$$

where $\tau_0$ is the total lifetime at sufficiently low Yb concentrations, and $N_c$ is the critical concentration. In silica, the upper laser state of $Yb^{3+}$ is almost purely radiative and $\tau_0$ is essentially equal to the radiative-relaxation time constant $t_{rad}$. The fiber cools when the term in parenthesis in (1) is negative. This occurs when the pump photon energy is lower than the average energy of the fluorescence photons *and* quenching is negligible ($\tau(N_0) \approx t_{rad}$). From (1), maximum ASF cooling is achieved with a large doped area, a short radiative lifetime, and a large energy difference between the pump and the fluorescence photons. There is also an optimum Yb concentration that maximizes the extracted heat, due to two competing effects – increasing $N_0$ increasing the number of heat engines (greater cooling), but it also decreases $\tau(N_0)$ (see (2)) (less cooling). Since the optimum Yb concentration increases with $N_c$, for efficient cooling it is paramount to develop glass with compositions and thermal histories that result in high $N_c$ values.

With these considerations in mind, the silica fiber used in this work was drawn from the same preform as the best performing fiber reported in [25]. It has a large core (21-μm diameter) co-doped with 2.0 wt.% Al to reduce concentration quenching [27] and increase $N_c$. This composition allowed the core to be highly doped with Yb (2.52 wt.%, or $1.93 \times 10^{26}$ Yb ions/m$^3$). From the measured emission spectrum of the fiber, the mean fluorescence wavelength was found to be relatively low (1003.9 nm), corresponding to a large average energy difference between the fluorescence and the 1040-nm pump photons (see Eq. (1)). The relatively short upper-state lifetime of the Yb ions (765 μs) is also beneficial to cooling. The preform was fabricated using conventional MCVD, followed by drawing at about 2000°C. With a numerical aperture of 0.13, the fiber is slightly multimoded ($V$ = 8.4 at 1064 nm). These measured parameter values, along with the parameter values inferred from absorption and temperature measurements (see *Supplemental Material*), are summarized in Table 1.

TABLE I
Yb-Doped Fiber Parameters

| Fiber Parameter | Parameter value |
| --- | --- |
| $Yb^{3+}$ concentration ($N_0$) | 2.52 wt.% Yb |
| $Al^-$ concentration ($N_{Al}$) | 2.00 wt.% Al |
| Core diameter ($2a$) | 21 μm |
| Numerical aperture ($NA$) | 0.13 |
| Radiative lifetime ($\tau_{rad}$) | 765 μs |
| Mean fluorescence wavelength ($<\lambda_f>$) | 1003.9 nm |
| Quenching lifetime ($\tau_q$) | 38 ms |
| Critical quenching concentration ($N_c$) | 21.0 wt.% Yb |
| Absorptive loss ($\alpha_{ba}$) | 18 dB/km |

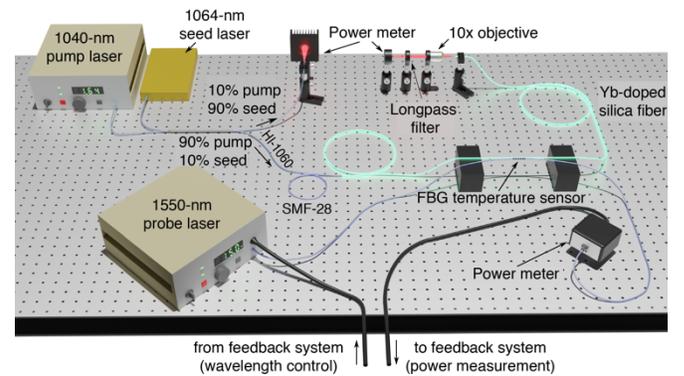

FIG. 1. Schematic of the experimental setup used to measure temperature changes induced in the Yb-doped silica fiber amplifier.

The fiber was core-pumped at 1040-nm to create both cooling and gain at 1064 nm (Fig. 1). A 90/10 fiber splitter combined the pump with the 1064-nm seed. The splitter output that carried 90% of the pump power and 10% of the seed power was spliced to an SMF-28 fiber, which was spliced to the input of the Yb-doped silica fiber. The SMF-28 fiber increased the optical transmission from the HI-1060 splitter fiber (6-μm-diameter core) to the gain fiber (21-μm diameter), since its intermediate core size (9-μm diameter) allowed for a more gradual evolution of the pump mode. This also reduced the fraction of pump power launched into the cladding of the Yb-doped fiber, thereby minimizing the heat generated by pump absorption by impurities in the cladding and jacket. The other splitter output was used as a tap to measure the power launched into the fiber amplifier. At the amplifier output, a long-pass filter removed the residual pump power and passed the signal to a power meter.

The induced temperature change was measured at seven locations along the Yb-doped fiber using a slow-light FBG sensor and interrogation technique described extensively in [28]. At each measurement location, a ~10-cm length of the Yb-doped fiber was stripped of its jacket to prevent reabsorption of the radially escaping ASF. The stripped section was placed in contact with the FBG sensor (see Fig. 1). A small amount of isopropanol was applied between the fibers to hold them together through capillary forces. When the Yb-doped fiber is pumped, its temperature changes and the two fibers quickly (a few seconds) reach thermal equilibrium [5].

Figure 2 shows an exemplary temperature measurement of the fiber amplifier. At time $t = 0$ s, the pump was abruptly turned on, causing the fiber to cool to ~130 mK below room temperature. After 15 s, the seed was also turned on, which induced a slight heating of the fiber (~20 mK) to a new steady-state value of ~110 mK below room temperature. The temperature change of the fiber amplifier was defined as the difference between the average temperature in the first 5 s when both the pump and signal were off and the average temperature in the last 5 s after both had been turned on. Each measurement was repeated three times and averaged.

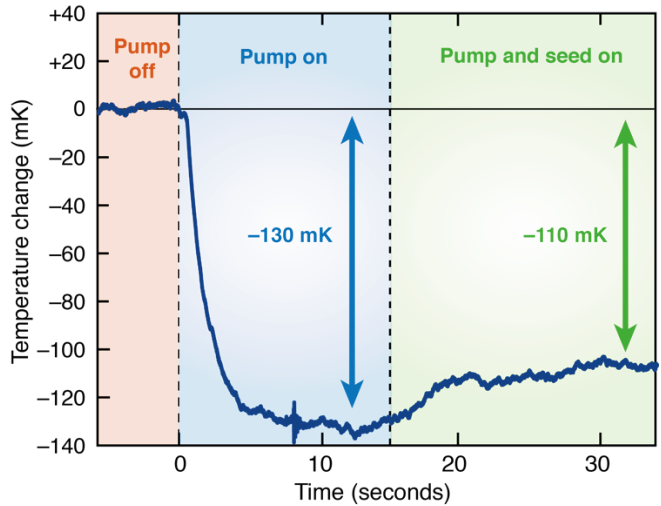

FIG. 2. Measured temporal trace of the temperature change in the Yb-doped silica fiber as the 1040-nm pump and 1064-nm seed are sequentially turned on, launching 1.64 W and 3 mW in the fiber core, respectively.

The first cooled amplifier consisted of a 2.74-m section of Yb-doped silica fiber core-pumped with 1.12 W at 1040 nm and seeded with 3 mW at 1064 nm. The temperature changes were measured to be negative at all seven locations (blue squares in Fig. 3) and the small-signal gain was 5.7 dB (11.7 mW of output power at 1064 nm). At low pump powers (below saturation), the Yb ions are insufficiently excited, resulting in low ASF and low extracted heat. At large pump powers (well above saturation), excitation of the Yb ions is saturated and the cooling rate is maximum. While the pump power in excess of the saturation power contributes very little to cooling, it is still absorbed by impurities. This process is essentially unsaturable and increases heating in proportion to this additional pump power. As a result of these two opposing effects, there is an optimum pump power that maximizes the extracted heat per unit length. Simulations using a model of ASF cooling in a fiber [19] predict that the optimum pump power for this fiber is 510 mW. In this experiment, the input pump power was 1.12 W and the residual pump power at the output of the amplifier was only 80.2 mW. It follows that (1) the negative temperature changes are smaller near both ends, where the pump power is either above (at the input) or below (at the output) this optimum value (510 mW), and (2) the lowest temperature is observed in the middle of the fiber, where the pump power is close to the optimum.

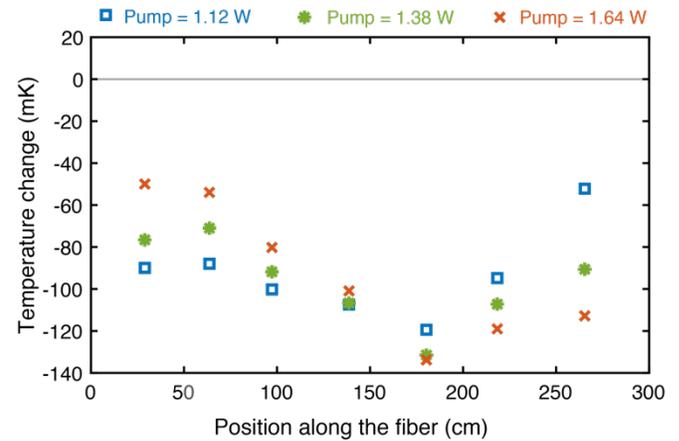

FIG. 3. Average measured temperature change ($n = 3$) at seven locations along a 2.74-m silica fiber amplifier for three different pump powers at 1040 nm. The fiber is core-pumped to create gain at 1064 nm for the 3-mW seed.

When the launched pump power was increased to 1.38 W, the signal output increased to 26 mW (a gain of 9.1 dB). As expected, the negative temperature change (green asterisks in Fig. 3) at the start of the fiber was now smaller than in the first case (blue squares) because the input pump power was further above the 510-mW optimum. This increase in launched power also resulted in a higher pump power at the fiber output end (154 mW). Since this value was now closer to the optimum, the output end of the fiber experienced a larger negative temperature change. The trend continued when the pump power was further increased to 1.64 W (red crosses in Fig. 3): cooling decreased near the input end and increased near the output end, where the residual pump power was now 247 mW. The signal amplification also increased, resulting in 11.4 dB of gain (44.6 mW of output power).

To determine the degree of uncertainty for the measured temperatures in Fig. 3, Fig. 4 (a-c) plots the three data points measured at each location. The solid curves are simulation results from the model of a radiation-balanced fiber laser described in [14], but with the cavity removed to simulate a fiber amplifier. All the fiber parameter values needed for these simulations were either measured or inferred from fits to independent cooling and absorption measurements (see *Supplemental Material*). Thanks to this comprehensive characterization, no fitting parameters were needed to generate the solid curves in Fig. 4. For each pump power, the measured temperatures agree well with simulations. From the model curves, the average temperature change is -107 mK for 1.12 W of input pump power, -105 mK for 1.38 W, and -93 mK for 1.64 W. The blue curve in Fig. 5 shows the simulated average temperature change as a function of input pump power. As expected, the cooling initially increases with pump power as more of the Yb ions are excited, but eventually plateaus and starts to decrease as the cooling saturates and the additional power induces more heating (around 1.1 W).

To increase the gain, the pump power was increased to 2.62 W and the length of the amplifier was increased to 4.35 m, the calculated optimum value for this pump power. This resulted in 16.9 dB of gain (or 146 mW of output power at 1064 nm) while the average temperature change along the amplifier length remained negative (see Fig. 4d). The higher pump power resulted in a positive temperature change at the input end, but once the power was sufficiently attenuated (after ~100 cm) the fiber cooled below room temperature for the remaining length. The solid curve in Fig. 4d is the temperature profile predicted by the amplifier model. It is in excellent agreement with the measured temperature changes. The average temperature of the amplifier was -24.4 mK.

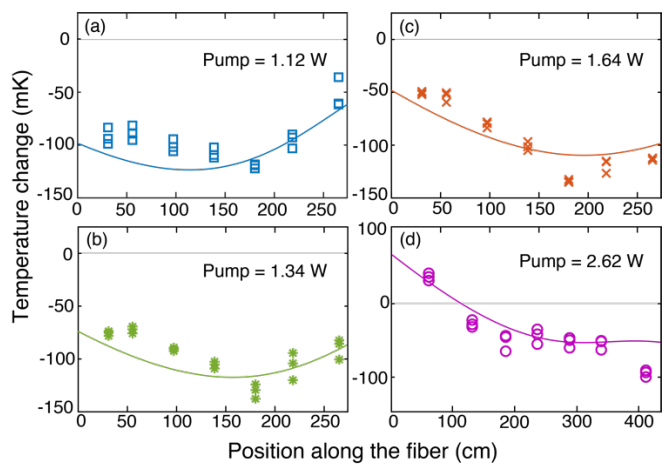

FIG. 4. Measured temperature change versus position along the fiber amplifier, and simulated dependencies using the model based on [14] for (a)-(c) a 2.74-m and (d) a 4.35-m amplifier fiber.

For each pump power, the gain of the 2.74-m amplifier was measured three times and plotted as a function of input pump power (red crosses in Fig. 5). The solid red curve was generated with the same model as above. The excellent fit confirms the accuracy of the model and of the measured and inferred parameter values for the Yb-doped silica fiber.

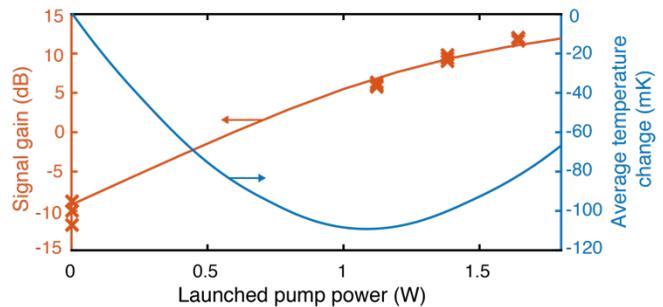

FIG. 5 Measured (red crosses) and simulated (red curve) small-signal gain at 1064 nm for a 3-mW seed as a function of input pump power at 1040 nm into a 2.74-m silica fiber amplifier, and the associated temperature change along the length of the amplifier as predicted by the model based on [14] (solid blue curve).

This work reports the first radiation-balanced fiber amplifier, in a fiber made not of a fluoride glass, but of silica, a significantly more ubiquitous, yet challenging material due to its historically low threshold for concentration quenching. The fiber used in this work has an aluminosilicate composition tailored to be doped with as much as 2,500 ppm of $Yb^{3+}$ while exhibiting negligible concentration quenching. The fiber was core-pumped at 1040-nm to cool the core while amplifying a 3-mW seed at 1064 nm. With 2.62 W of launched pump power, the 4.35-m fiber amplifier produced 16.9 dB of small-signal gain while maintaining a negative average temperature of -24.4 mK below room temperature. A model of a radiation-balanced fiber amplifier was developed to validate the gain and temperature profile recorded for various pump powers. This work establishes that fiber technology is now capable, for the first time in history, to produce silica fibers with such chemical purity and low degree of quenching that they can be doped with unprecedented concentrations of ytterbium. In such a fiber, we showed that light can be coherently amplified with a gain approaching 20 dB and generate *no net internal heating*. This fundamental development is ushering silica fibers into a new era where it is possible to create fiber lasers and amplifiers with groundbreaking coherence and stability.

The authors would like to thank Carston Langrock (Stanford University) for the generous amount of equipment he has lent to this project, and Nanjie Yu for measuring the fiber's radiative lifetime. This work was supported by the Air Force Office of Scientific Research grant number #FA9550-16-1-0383.

See *Supplemental Material* for supporting content.

~~~~~~~~~~

# Supplementary material for "A Radiation-Balanced Silica Fiber Amplifier"

Jennifer M. Knall,[1,*] Magnus Engholm,[2] Tommy Boilard,[3] Martin Bernier,[3] and Michel J. F. Digonnet[1]

[1]Edward L. Ginzton Laboratory, Stanford University, Stanford, California 94305, USA
[2]Division of Electronics Design, Mid Sweden University, SE-85170 Sundsvall, Sweden
[3]Centre d'optique, photonique et laser (COPL), Université Laval, Québec, Canada QC G1V 0A6

## Fiber Characterization

### 1. Upper-state lifetime and emission cross-sections

The upper-state lifetime of the Yb ions in the silica fiber was measured by end-pumping a 1-mm piece of fiber with a 976-nm pulsed semiconductor laser (pulse width of 10 µs and pulse period of 6.5 ms). An avalanche photodiode was used to measure the decay of the fluorescence power at the output of the segment. This resulted in an exponential decay with a single slope on the log-log scale, confirming the absence of significant non-radiative relaxation. The fluorescence spectrum was measured by pumping the same sample continuously, collecting the fluorescence emitted from the side of the fiber with a multimode fiber, and measuring the spectrum of the collected light with a grating-based optical spectrum analyzer. The emission cross-section spectrum was then calculated from the spontaneous emission spectral intensity using the well-known Füchtbauer-Ladenburg equation [1].

### 2. Fiber composition

The composition of the fiber was measured using energy-dispersive X-ray spectroscopy (EDX) on a Hitachi SU5000 high-resolution scanning electron microscope (HRSEM) in variable pressure mode with 50 Pa of pressure, 15 keV of accelerating voltage, and a 10-mm working distance. Using the EDX data, the Al and Yb concentrations were averaged across the area of the core in the region extending to a radius of about 11 µm, giving 2.00, and 2.52 wt.%, respectively. The average $Yb^{3+}$ number density, therefore, was determined to be $1.93 \times 10^{26}$ Yb ions/m$^3$.

### 3. Small-signal absorption and saturation power

Cut-back measurements were performed to obtain precise absorption values at 1040 nm and 1064 nm, which were needed to carry out all simulations. Since the Yb-doped silica fiber is slightly multimoded, the absorption parameter at each wavelength depends on the launching conditions. Therefore, to ensure that the appropriate values were obtained, the absorption measurements were performed with the same experimental set-up as the one illustrated in Fig. 1 of the main text. For each wavelength, the power was varied and the corresponding output power after 2.74 m of Yb-doped silica fiber was measured with a thermal power meter. The fiber was then cut so that only 10 cm remained spliced to the output of the SMF-28 fiber after the 90/10 splitter. The power sweep was repeated and the output power remeasured. The red crosses in Fig. S1 represent the output power $P_{out}$ measured before the cut-back versus the input power $P_{in}$ measured after the cut-back. This dependency was then fitted to the following absorption equation, which assumes a top-hat mode distribution for the power in the fiber:

$$P_{out} = P_{in} exp\left(\frac{P_{in}-P_{out}}{P_{sat}} - \alpha_0 L\right) \quad (S1)$$

where $L$ = 2.64 m is the length of fiber removed by cleaving. With $\alpha_0$ and $P_{sat}$ as fitting parameters, (S1) was fitted to minimize the $\chi^2$ between the measured and predicted output power dependencies (blue curves in Fig. S1). For the 1040-nm pump, this resulted in $\alpha_{0,1040}$ = 2.0 m$^{-1}$ and $P_{sat,1040}$ = 452 mW. At 1064 nm, there was not enough power to reach $P_{sat,1064}$ (calculated from measured spectroscopic parameters to be ~420 mW). In this low-power regime, the first term in the exponential in (S1) approaches zero and the only fitting parameter left is $\alpha_{0,1064}$, which is all the data needed for this work since the power at 1064-nm is maintained well below the saturation power. Fitting the small-signal absorption for the 1064-nm signal resulted in $\alpha_{0,1064}$ = 0.83 m$^{-1}$.

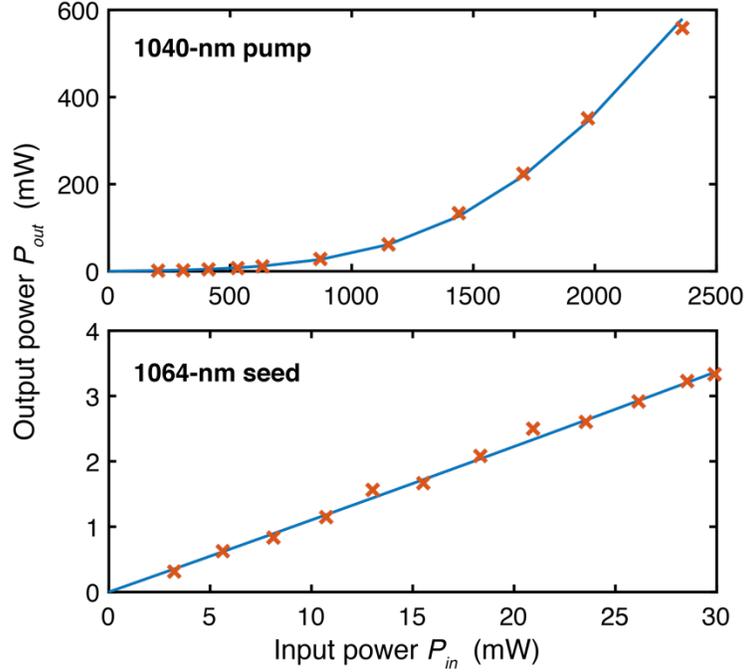

FIG. S1. Results from cut-back measurements performed on the Yb-doped fiber at 1040 nm and 1064 nm. The measured output power $P_{out}$ is plotted as a function of pump power $P_{in}$ launched into the fiber for each wavelength, and the data points are fitted to a model of saturated absorption in a fiber.

## 4. Absorptive loss and critical quenching concentration

Temperature measurements were performed to infer the residual absorptive loss $\alpha_{ba}$ due to impurities and the critical quenching concentration $N_c$ of the Yb-doped fiber. The same experimental set-up as illustrated in Fig. 1 of the main text was used. The FBG temperature sensor was placed in contact with a stripped section of the Yb-doped silica fiber 60-cm from the input end, and a small amount of alcohol was applied between the fibers for better thermal contact. The doped fiber was core-pumped at 1040 nm and the induced temperature change was measured at the FBG location for various powers up to 1.8 W (red crosses in Fig. S2). The data was then fitted to a published model of ASF cooling in a fiber [2] to minimize the $\chi^2$ between the measured and predicted pump power dependencies (solid blue curve in Fig. S2). Since the pump absorption coefficients were independently measured, the only fitting parameters were the critical quenching concentration and the absorptive loss. These parameter values were inferred to be $N_c$ = 21.0 wt.% Yb (1.63x10$^{27}$ Yb$^{3+}$/m$^3$) and $\alpha_{ba}$ = 18 dB/km, which are comparable to the values reported for the highest cooling fiber in [3] drawn from the same preform. The high critical quenching concentration in these fibers is more than a factor of 15 greater than in the best commercial Yb-doped silica fibers [4], which equates to an $N_c$ that is almost half as large as in Yb-doped ZBLAN fibers (3.54x10$^{27}$ Yb$^{3+}$/m$^3$) [5]. This is fundamental to the exceptional cooling performance of these silica fibers. This work establishes a silica composition that can be doped with Yb at a concentration level approaching that of ZBLAN fibers, offering a new and much more versatile medium to produce a wide range of short, efficient, optically cooled fiber lasers.

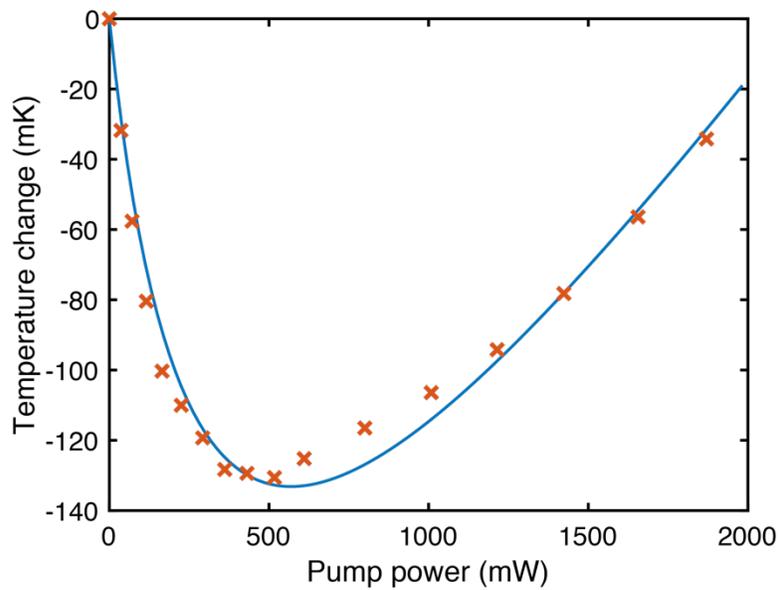

FIG. S2. Temperature change as a function of 1040-nm pump power at the measurement location for the Yb-doped silica fiber presented in this work. The data is fit to a model of ASF cooling in a fiber [2] to infer the absorptive loss and critical quenching concentration.